\documentclass[12pt]{article}

\usepackage{amsmath,amsfonts,amssymb}

\def\1barra{1\! \hskip -1.1pt {\rm l}}

\def\0barra{{\rm O} \!\hskip -3.7pt {\rm l} }

\title{Imaginary Phases in Two-Level Model with Spontaneous Decay.}

\author{ A.C. Aguiar Pinto$^{(1,2)}$ and M.T. Thomaz$^{(1)}$
 \\
\\
\baselineskip =10pt
{ \small \it $^{(1)}$ Instituto de F\'\i sica - Univ. Federal
                      Fluminense
\vspace{-0.2cm}}\\
{\small \it Av. Gal. Milton Tavares de Souza s/n.$\!\!^\circ$,
                \vspace{-0.2cm} }\\
{ \small \it CEP: 24210-340, Niter\'oi, R.J.,  Brazil }  \\
{\small \it $^{(2)}$ Instituto de F\'{\i}sica - Univ. Federal do Rio de Janeiro}
\vspace{-0.2cm}\\
\vspace{-0.2cm}{\small \it Caixa Postal: 68528, }\\
{\small \it CEP: 21945-970, Rio de Janeiro, R.J., Brazil}\\}

\begin{document}

\maketitle

\begin{abstract}


We study a two-level model coupled
to the electromagnetic vacuum and to an external classic electric field
with fixed frequency.  The amplitude of the external electric field
is supposed to vary very slow in time.
Garrison and Wright [{\it Phys. Lett.} {\bf A128} (1988) 177]
used the non-hermitian Hamiltonian approach to study the
adiabatic limit of this model and obtained that the
probability  of this two-level system to be in its upper level
has an imaginary geometric phase. Using the master equation for describing
the time evolution of the two-level system we obtain that
the imaginary phase due to dissipative effects is time dependent, in opposition
to Garrison and Wright result. The present results show that the non-hermitian
hamiltonian method should not be used to discuss the nature 
of the imaginary phases in open systems.

\end{abstract}

\vfill

\baselineskip=12pt



\newpage

\baselineskip=18pt

\section{Introduction}

The natural approach to study quantum systems in contact with an
environment is the density operator method. It allows one to study
quantum states evolving from pure to mixed states. In general,
exponential decay in the time evolution of the population at each
excited state of the quantum system occurs due to the exchange between
the system and its environment. An alternative approach to obtain
those exponential decays in the probability  is through complex
energies obtained from a phenomenological non-hermitian
Hamiltonian\cite{moiseyev98,baker83,baker84,dattoli90b}. The
non-hermitian Hamiltonian method has been very fruitful in approaching
various physical problems as, for example, the multiphoton
ionization\cite{baker83,baker84,dattoli90b,themelis01} and
free-electron laser theory\cite{dattoli88a,dattoli88b}. As mentioned
by Baker in reference \cite{baker84}, it is expected that the
non-hermitian Schr\"odinger equation be a {\it bona fide} description
of the interaction among the parts of a system when the intervals of
time are sufficiently short such that the coherence state of a
subsystem is not destroyed by its interaction with the environment.


Since the disclosure of geometrical phases by M.V. Berry in
1984\cite{berry84} in cyclic Hamiltonians evolving adiabaticaly, there
has been significant search for geometrical phases in other physical
contexts. For example, Joye {\it et al}.\cite{joye91} and
Berry\cite{berry90} independently showed that the transition
probability  of instantaneous eigenstates of non-real Hamiltonians in
the non-adiabatic regime gets an imaginary geometric phase. This
imaginary geometric phase was measured by Zwanziger {\it et al}. in a
two-level system\cite{zwanziger91}. A great part of the work done on
geometric phases has been on pure states. Since the work by Uhlmann in
1986\cite{uhlmann86,uhlmann89}, however, the study of holonomy has been
extended to mixed states under unitary
evolution\cite{dabrowski1,dabrowski2,anadan}. More recently, Ericsson
{\it et al}.\cite{ericsson} obtained the expression of the geometric
phase of a quantum system interacting with its environment when 
the unitary evolution of the whole system (including the
environment) is known. Certainly the discussion of holonomy in mixed states is
a very interesting point, since the correct expression of the
geometric phase for mixed states under unitary evolution is still
under debate, as in a very recent work by Singh {\it et
al}.\cite{singh}. However, that is not the issue of the present
communication.  Another equally interesting question  about the
imaginary phases in transition probabilities of open systems
is the correctness of the application of the non-hermitian hamiltonian
approach\cite{garrison88,dattoli90,mostafazadeh,korsch} to
discuss the nature of such phases.

In particular Garrison and Wright\cite{garrison88} used the
non-hermitian Hamiltonian method to study a two-level model with
linewidths in the presence of an external electric field, obtaining
the adiabatic limit of the probability  of this system being in its
upper level, after the external field has returned to its original
configuration. They concluded that the decaying factor has an imaginary
correction to Berry's phases. Their result for the 
two-level model coupled to an external classical electromagnetic
field with spontaneous emissions is already contained in the phases
(75) and (76) of reference \cite{baker84} for any open system
described by a non-hermitian Hamiltonian. It is the mathematical
structure of these phases in the non-hermitian Hamiltonian approach
that makes the imaginary phase derived by Garrison and Wright
in reference \cite{garrison88} to have an imaginary geometric contribution.

Garrison and Wright mentioned in their conclusion of reference
\cite{garrison88} that the nature (time-dependent or path-dependent)
of their imaginary phase due to dissipation effects should be
reexamined using the density matrix approach; this is the aim of the
present letter.

In reference \cite{physa2002} we studied the adiabatic
 limit of any periodic non-degenerate
Hamiltonian using the density matrix approach (extending the
discussion carried out by Born and Fock in reference \cite{born28} to
the density matrix in the basis of instantaneous eigenstates of the Hamiltonian),
and we concluded that for a quantum system to get an
imaginary correction to Berry's phases in dissipative phenomena,
the functions in the integrals of the decay exponentials in
the entries of the density matrix  would have to satisfy
special conditions. Let $e^{-\int_0^t dt^\prime c(t^\prime)}$ be a typical
decreasing exponential originated by the
presence of the dissipative effects. For the integral in the
exponential be written as a path-dependent integral, the function $c(t)$
must have the form

\begin{equation}
c(t) = \varphi_i(t)\frac{d}{dt}\Big(\Psi_i(t)\Big),
\label{44.4}
\end{equation}

\noindent and the functions $\Psi_i(t)$ have to satisfy two
conditions: $i)$ they must not be explicitly time-dependent; $ii)$ the
time-dependence of functions $\Psi_i(t)$ must come only from their
dependence on the set of parameters $\vec{{\bf k}} (t)\equiv (
k_1(t),k_2(t), \cdots, k_l(t))$. We point out that there is no
restriction to the regime of the time variation of the set of
parameters $\vec{\bf k}(t)$ and that it has not to be a periodic
function in time, i.e., the path in the $\vec{\bf k}$-parameter space
has not to be closed (see details in reference \cite{physa2002}).


\vspace{.5cm}


\begin{subequations} \label{gar1}

In reference \cite{garrison88}, Garrison and Wright considered a
two-level system interacting with a classic external
electromagnetic field

\begin{equation}
\vec{E}_{clas}(t) = {\rm{\bf Re}}\Big[\vec{e}\;{\cal E}(t) e^{i \nu t}
\Big],  \label{2.4}
\end{equation}

\noindent where the amplitude ${\cal{E}}(t)$ varies
very slowly. The two energy states were
supposed to have a linewidth. They used the
time-dependent non-hermitian Schr\"odinger (Bethe-Lamb) equation
in the rotating wave approximation (RWA)\cite{garrison88,lamb87}

\begin{eqnarray}
i \frac{d}{dt} \left( \begin{array}{c}
C_a (t) \\ C_b (t) \end{array}\right) =
\left( \begin{array}{cc}
-\frac{1}{2}i\gamma_a & V^*e^{i\Delta t} \\
V e^{-i\Delta t} & -\frac{1}{2}i\gamma_b \end{array}\right)
\left( \begin{array}{c}
C_a(t) \\ C_b(t) \end{array}\right),  \label{gar1.1}
\end{eqnarray}

\noindent where $\gamma_a$ and $\gamma_b$ are the decay rates for
the upper and lower levels, respectively, 
and $C_a (t)$ and $C_b (t)$ are the corresponding probability
amplitude  of states $|a\rangle$ and $|b\rangle$.
Moreover, 
$\Delta=\omega-\nu$, $\omega=(E_a-E_b)/\hbar$,
$V= \vec{\mu} \cdot \vec{e}\;{\cal E}(t)/2\hbar$ and
 $\vec{\mu}= \vec{\mu}_{ab}$ is the  electric
dipole matrix element. Letting ${\cal E}(t) = {\cal E}_0 e^{i\phi(t)}$,
where $\phi(0)=0$ and $\phi(T)=2\pi$, with $T\gg \frac{2\pi \hbar}{E_a-E_b}$
they obtained a complex Berry's phase $\beta_-$ in the
expression for the probability  of the
quantum system to be in the state $a$ at time $T$,

\begin{equation}
\beta_- =\frac{1}{2}[2\pi(1-\cos(\theta_0))] =
\frac{1}{2}[2\pi(1-\frac{\Delta - i\delta}{\sqrt{|2V_0|^2 +
(\Delta - i\delta)^2}})] \label{gar1.2}
\end{equation}

\noindent where $\delta \equiv (\gamma_a-\gamma_b)/2$ and
$V_0=  \vec{\mu}\cdot \vec{e}\; {\cal E}_0 /2\hbar$.

\end{subequations}

\vspace{.5cm}

In the present work we want to verify if the result about the
path-dependence of the imaginary phases derived by non-hermitian
method is faithful. To do so, we apply the density matrix approach
presented in reference \cite{physa2002} to study the same physical
problem considered in reference \cite{garrison88} with $\gamma_b=0$.
We wish to compare the nature (path or time-dependent) of the
imaginary phases in the expression of the probability  of the atomic
electron, initially at state $|a\rangle$, being at the same state at
time $t=T$, derived by the density matrix approach with the one
obtained  in the  non-hermitian phenomenological hamiltonian framework.

\vspace{0.5cm}

Let the two-level system that describes an atomic electron be
represented by the Hamiltonian ${\bf H}_e$. The atomic electron interacts
with a classic external electromagnetic field $\vec{E}_{clas}(t)$ (see
eq.(\ref{2.4})) and with  the electromagnetic vacuum (the vacuum electromagnetic field
operator being represented by ${\vec{\bf E}}_0(\vec{x})$). The total
Hamiltonian for this model in the Schr\"odinger picture is\cite{louisell73}

\begin{equation}
{\bf H}_T = {\bf H}_e + {\bf H}_f + {\bf H}_{int},
\label{1.1}
\end{equation}

\noindent with

\begin{subequations} \label{2}

\begin{eqnarray}
{\bf H}_e &=& \frac{{\bf \vec{P}}^2}{2m} + V({{r}}), \label{2.1} \\
{\bf H}_f &=& \sum_{\vec{k}}\sum_{\lambda =1}^2
\hbar \omega_{\vec{k}} \Big({\bf a}^\dagger_{\vec{k}\lambda}
{\bf a}_{\vec{k}\lambda} + \frac{1}{2} \Big), \label{2.2}
\end{eqnarray}

\vspace{-.3cm}

\noindent and

\vspace{-.3cm}

\begin{eqnarray}
{\bf H}_{int} &=& -e\Big( \vec{E}_{clas}(t) + {\vec{\bf E}}_0(\vec{x}) \Big) \cdot
{\bf \vec{r}}, \label{2.3}
\end{eqnarray}

\noindent where ${\bf \vec{P}}$ is the momentum operator associated to the
atomic electron, $ V({{r}})$ is the spherical interaction potential between
the electron and the rest of atom and ${\bf H}_f$ is the
Hamiltonian of the electromagnetic energy operator of the electromagnetic vacuum.
The operator ${\bf a}^\dagger_{\vec{k}\lambda}$ (${\bf a}_{\vec{k}\lambda}$)
creates (destroys) a photon with momentum $\vec{k}$ in the polarization
state $\lambda$. We also have: $\omega_{\vec{k}} = c |{\vec{k}}|$.

Following reference \cite{garrison88}, the
classic external electromagnetic field $\vec{E}_{clas}(t)$ is given by eq.(\ref{2.4})
 assuming that ${\cal{E}}(t)$ varies very
slowly and $\vec{e}$ is constant. The electromagnetic field operators
in the vacuum ${\vec{\bf E}}_0(\vec{x})$ in the Schr\"odinger picture have the
expansion

\begin{equation}
{\vec{\bf E}}_0 (\vec{x}) = \sum_{\vec{k}}\sum_{\lambda =1}^2
\vec{\epsilon}_{\vec{k}\lambda} \sqrt{\frac{\hbar\omega_{\vec{k}}}{2
\epsilon_0V}} \Big(-i{\bf a}^\dagger_{\vec{k}\lambda}e^{-i\vec{k}\cdot\vec{x}} +
i{\bf a}_{\vec{k}\lambda}e^{i\vec{k}\cdot\vec{x}} \Big). \label{2.5}
\end{equation}

\end{subequations}

The dynamics of the density operator of the
complete system is driven by the Liouville von-Neumann equation. Taking the
trace over the electromagnetic degrees of freedom we obtain the
master equation for the reduced density matrix of the atomic electron
$\bar\rho(t)$ written in the basis of the eigenstates of ${\bf H}_e$
(${\bf H}_e | i \rangle = E_i | i \rangle  $). The dynamics of $\bar\rho(t)$
in the electric dipole approximation and in the RWA is\cite{agarwal73,carmichael93}

\begin{eqnarray}
\frac{d}{dt}\bar\rho(t) &=& \frac{1}{i}[(\omega_0+\Omega_+)\sigma_z, \bar\rho(t)] -
\frac{1}{i\hbar}\vec{\mu}_{ab}\cdot \vec{E}_{clas}(t) [\sigma_x , \Lambda^0(t)]
 \nonumber \\
%
%
  && \hspace{-1.5cm} + \gamma \Big(2\sigma_- \bar\rho(t) \sigma_+
  -\{\sigma_+\sigma_-,\bar\rho(t)\} \Big)
+ \frac{\vec{\mu}_{ab}\cdot \vec{E}_{clas}(t)}{\hbar^2}
 \Big( -2A(t)\bar\rho(t)    \nonumber \\
%
%
 && \hspace{-1.5cm}+  2 A(t)\sigma_x \bar\rho(t) \sigma_x
+ B(t)\big[[\sigma_y, \bar\rho(t)], \sigma_x\big] \Big), \label{3}
\end{eqnarray}

\noindent where $\sigma_z$ and $\sigma_{\pm}$ are the Pauli matrices with
$\sigma_{\pm}= \frac{1}{2}(\sigma_x \pm i \sigma_y)$,
$ \vec{\mu}_{ab}$  is the electric dipole matrix
(we are supposing $\vec{\mu}_{ab}$ to be real) with
$\vec{\mu}_{ab} \equiv e \langle a | {\bf{\vec{r}}}|b \rangle$,$(e>0)$,

\begin{subequations}

\begin{equation}
\gamma= \frac{\pi}{\hbar^2}
\sum_{\lambda=1}^2\int d^3\vec{k} \; \eta({\vec{k}}) \;
|g(\vec{k},\lambda)|^2 \;\delta(\omega_0-\omega_{\vec{k}}), \label{gamma}
\end{equation}

\noindent being $g(\vec{k},\lambda)= -i
\vec{\epsilon}(\vec{k},\lambda)\cdot \vec{\mu}_{ab}
\sqrt{\frac{\hbar\omega_{\vec{k}}}{2\epsilon_0V}}$ and
$\eta({\vec{k}})$ is the density of states introduced in
the integration,

\begin{equation}
\omega_0 \equiv \frac{E_a- E_b}{2\hbar} = \frac{\omega}{2},
\end{equation}

\begin{equation}
\Lambda^0_{ij}(t) \equiv e^{-\frac{i}{\hbar} (E_i-E_j)t}\rho_{ij}(0),
     \hspace{1cm} i, j = a, b,
\end{equation}

\noindent and

\begin{equation}
\Omega_+ = - \frac{\gamma}{\pi} \ln \Big[ \Big| \frac{\omega_c}{\omega_0}
- 1 \Big| \Big( \frac{\omega_c}{\omega_0} + 1 \Big)\Big],    \label{Omega+}
\end{equation}

\noindent where $\omega_c$ is the cutoff frequency that preserves the
dipole approximation ( $\omega_c < c/a_0$, with $a_0$ being the
atomic Bohr radius). The cutoff $\omega_c$ becomes a parameter
of the effective model to be determined by fitting to experimental data.
The term $\Omega_+$ corresponds to the frequency shift\cite{agarwal73}.
The elements $\rho_{ij} (0)$ are the initial  values of the entries
of matrix $\bar{\rho}(t)$.
The functions $A(t)$ and $B(t)$ appearing in eq.(\ref{3}) are defined as:
$A(t) \equiv \int_0^t dt^\prime \vec{\mu}_{ab}\cdot \vec{E}_{clas}(t^\prime)
\cos[\frac{(E_a - E_b)}{\hbar}(t-t^\prime)]$   and
$B(t) \equiv \int_0^t dt^\prime \vec{\mu}_{ab}\cdot \vec{E}_{clas}(t^\prime)
\sin[\frac{(E_a - E_b)}{\hbar}(t-t^\prime)]$.

\end{subequations}


We remind that the density matrix of a two-level model
must satisfy two conditions: {\it i}) $Tr(\bar\rho (t)) =1$  $\;$ and
{\it ii}) $\rho_{ba}(t)= (\rho_{ab}(t) )^* $, where $\rho_{ab}(t)
\equiv \langle a | \bar\rho(t) | b \rangle$. As a consequence of those
conditions, the density matrix has only two independent elements; we 
choose $\rho_{aa}(t)$ and $\rho_{ab}(t)$ to be such elements.
>From eq.(\ref{3}) the time equations for those two elements are

\begin{subequations} \label{4}

\begin{eqnarray}
\frac{d}{dt} \rho_{aa}(t) &=&  \Big[- 2\gamma
-4 \vec{\mu}_{ab}\cdot \vec{E}_{clas}(t)
     {\rm{\bf Re}}\Big( e^{i(E_a-E_b)t} G(t)\Big)\Big]\rho_{aa}(t) \nonumber \\
%
%
&&  \nonumber \\
%
%
 && \hspace{-1.8cm} +  2\vec{\mu}_{ab} \cdot \vec{E}_{clas}(t)
{\rm{\bf Im}} \Big[ e^{-i(E_a-E_b)t} \rho_{ab}(0)\Big]
+ 2\vec{\mu}_{ab} \cdot \vec{E}_{clas}(t)
{\rm{\bf Re}}\Big[ e^{i(E_a-E_b)t} G(t)\Big]  \nonumber \\
%
                   \label{4.1}
\end{eqnarray}

\noindent and

\begin{eqnarray}
\frac{d}{dt} \rho_{ab}(t) &=&
 \Big(-2i (\omega_0 + \Omega_+) -\gamma
 -2\vec{\mu}_{ab} \cdot \vec{E}_{clas}(t) G(t) e^{i(E_a-E_b)t} \Big)\rho_{ab}(t)
     \nonumber \\
 %
 %
   \nonumber \\
 %
 %
 && \hspace{-1.5cm} + 2\vec{\mu}_{ab} \cdot \vec{E}_{clas}(t) G^*(t)
     e^{-i(E_a-E_b)t}\rho_{ba}(t)
+ i \vec{\mu}_{ab} \cdot \vec{E}_{clas}(t) ( 1 - 2 \rho_{aa}(0)).
                 \label{4.2}
\end{eqnarray}

\end{subequations}

\noindent The expression of the function $G(t)$ is:
$G(t) \equiv \int_0^t dt^\prime \vec{\mu}_{ab} \cdot
\vec{E}_{clas}(t^\prime)e^{-i(E_a-E_b)t^\prime}$.
 We see from eq.(\ref{4.2})  that the real and
imaginary parts of $\rho_{ab}(t)$ are coupled due to non-linear
effects in the classic external electromagnetic field.
 From eqs.(\ref{4}) on,  we will be  using natural units ($\hbar=c=1$).

In the same manner as in references \cite{physa2002,romero01}, in
order to study the $T\rightarrow \infty$ limit of eqs.(\ref{4}) we
apply the transformation

\begin{equation}
\tilde{\rho}_{ij}(t) \equiv e^{i(E_i-E_j)t}\rho_{ij}(t),
         \hspace{1cm} i,j = a,b
\end{equation}

\noindent and change the time scale to $s = t/T $; in the
limit $T\rightarrow \infty$, one obtains

\begin{subequations} \label{5}

\begin{eqnarray}
\frac{d}{ds} \tilde{\rho}_{aa}(s) &\approx&
 T \Big[ -2\gamma
  - 2\vec{\mu}_{ab}\cdot \vec{e}\, {\cal E}_0  \,
   {\rm {\bf Re}}(\tilde{G}(s) e^{-i(\phi(s)- \Delta Ts)}) \Big]
                  \tilde{\rho}_{aa}(s) \nonumber \\
%
%
 && \hspace{-1cm}
     +T \vec{\mu}_{ab}\cdot \vec{e}\, {\cal E}_0\,\Big[
 {\rm {\bf Im}}(e^{i(\phi(s)- \Delta Ts)} \tilde{\rho}_{ab}(0) )
 +{\rm {\bf Re}}(\tilde{G}(s) e^{-i(\phi(s)- \Delta Ts)})
        \Big]   \label{5.1}
\end{eqnarray}

\noindent and

\begin{eqnarray}
\frac{d}{ds} \tilde\rho_{ab}(s) &\approx& - \Big(2i\Omega_+
+ \gamma + \vec{\mu}_{ab}\cdot
\vec{e}\, {\cal E}_0 \tilde{G}(s) e^{-i(\phi(s) -\Delta Ts)}\Big)T
\tilde{\rho}_{ab}(s)  \nonumber \\
%
%
 && \hspace{-1.5cm} +\vec{\mu}_{ab}\cdot
\vec{e}\, {\cal E}_0 \, \tilde{G}^*(s) e^{-i(\phi(s) -\Delta Ts)}
 T \tilde{\rho}_{ba}(s)
 + i  T \vec{\mu}_{ab} \cdot \vec{e}\, \frac{{\cal E}_0}{2}
( 1 - 2 \tilde\rho_{aa}(0))e^{-i (\phi(s)- \Delta T s )} .  \nonumber \\
%
%
    \label{5.2} \end{eqnarray}

\end{subequations}

\noindent As we write eqs.(\ref{5}),  we are assuming that
 $\Delta \equiv 2\omega_0-\nu \sim
\frac{2\pi}{T}$. The terms proportional to
$e^{ \pm i ((2\omega_0 + \nu)T s + \phi(s))}$
do not contribute in the limit $T\rightarrow \infty$
(as shown in references
\cite{physa2002,romero01}) and the terms proportional to
 $ e^{ \pm i(\Delta T s - \phi(s))} $ contribute to the dynamics
 of the density matrix in the resonance region when $\Delta \sim
\frac{2\pi}{T}$. The function  $\tilde{G}(s)$ appearing in eqs.(\ref{5})
is given by
$\tilde{G}(s)= \vec{\mu}_{ab} \cdot\vec{e} \; \frac{{\cal E}_0 T}{2}
     \int_0^{sT}  ds^\prime  e^{i(\phi(s^\prime)- \Delta Ts^\prime)}$.

Equation (\ref{5.1}) gives us the probability that the atomic
electron be at state $|a\rangle$. Its solution after one period T is

\begin{eqnarray}
{\rho}_{aa}(T) &=& \Big\{ {\rho}_{aa}(0) +
i\int_0^T dt^\prime  i
\Big( \frac{d\tilde{G}(t^\prime)^*}{dt^\prime} \tilde{\rho}_{ba}(0)-
   \frac{d\tilde{G}(t^\prime)}{dt^\prime} \tilde{\rho}_{ab}(0)        \Big)
 e^{2(\gamma t^\prime +|\tilde{G}(t^\prime)|^2) }  \nonumber \\
%
%
   \nonumber \\
%
%
&+& \frac{1}{2} \int_0^T dt^\prime  e^{2\gamma t^\prime}
  \frac{d[e^{2|\tilde{G}(t^\prime)}|^2 ]}{dt^\prime}
   \Big\}\; e^{-2(\gamma T +|\tilde{G}(T)|^2)}\Big\}. 
                 \label{6.1}
\end{eqnarray}

\noindent Since the function $\tilde{G}(t)$ is an
explicit time-dependent function,  the condition {\it i} in
 eq.(\ref{44.4}) is not satisfied and
none of the integrals in eq.(\ref{6.1}) is a time-independent
integral.  We have an overall exponential decay  (imaginary factor),
but from  the definition of the constant $\gamma$ (see eq.(\ref{gamma})),
we obtain that the exponential
$e^{-2\gamma t}$ is not a geometric (path-dependent) imaginary phase,
but a time-dependent one as well as the contribution to the
decreasing exponential coming  from $|\tilde{G}(T)|^2$.

For the sake of completeness, we should also examine the solution of
eq.(\ref{5.2}) after a period T.
Differently from reference \cite{romero01}, here the dynamical equation of
$\rho_{ab}(t)$ couples  its real and imaginary parts under the regime of strong
classic external electric field.   We get a $SU(2)$ structure for the solution
of  $\rho_{ab}(t)$.   Calling $\rho_1(t)= \rho_{ab}(t) $  and  $\rho_2(t)=
\rho_{ba}(t) $, the   solution of eq.(\ref{5.2}) is

\begin{eqnarray}
 \rho_I (T) &=&\Big[ {\cal T} \Big( e^{\int_0^T  dt^\prime \;
    \vec{\cal B}(t^\prime)\cdot\vec{\sigma}}   \Big) \Big]_{IJ} 
\Big\{\int_0^T dt^\prime \;  \Big[ {\cal T} \Big( e^{\int_T^0
dt^{\prime\prime}\;  \vec{\cal B}(t^{\prime\prime})\cdot\vec{\sigma}} \Big)
\Big]_{JK} \times                   \nonumber  \\  %
 %
    \nonumber \\
%
%
 && \hspace{-2cm} 
        \times e^{(\gamma t^\prime +{\rm{\bf Re}}^2 (\tilde{G}(t^\prime)) 
                       -  {\rm{\bf Im}}^2 (\tilde{G}(t^\prime)))}  d_K(t^\prime)
					        + \rho_J(0)     \Big\} \;
  e^{(-1)^I  2i\omega_0 T  } 
 e^{-(\gamma T +{\rm{\bf Re}}^2 (\tilde{G}(T)) - {\rm{\bf Im}}^2
(\tilde{G}(T)))} ,      \label{7}
 \end{eqnarray}
  
\noindent with $I=1,2$ and $J,K=1,2$. In these two last indices we are using
the implicit sum notation on the r.h.s. of eq.(\ref{7}).  The symbol
${\cal T}$  means the time-ordering integrals\cite{time_operator}, and 
$\vec{\sigma}$ are the Pauli matrices. The elements of the  column $d_J(t)$
are: $d_1(t)= d_2^* (t) =  i \vec{\mu}_{ab}\cdot \vec{e}\, \frac{{\cal
E}_0}{2}[1-2\rho_{aa} (0)]          e^{-i(\phi(t) -\Delta t)}$. The components
of the vector  $\vec{\cal B}$ are:
${\cal B}_x (t) = \frac{d[|\tilde{G}(t)|^2]}{dt}$,
${\cal B}_y (t) =
 2 \Big[ {\rm {\bf Im}}(\tilde{G}(t)) \frac{d\big( {\rm {\bf Re}}(\tilde{G}(t))  \big)}{dt}    
 - {\rm {\bf Re}}(\tilde{G}(t)) \frac{d\big( {\rm {\bf Im}}(\tilde{G}(t))  \big)}{dt}   \Big]$
and 
${\cal B}_z (t) =
 -2i  \Big[  \Omega_+
+ \frac{ d}{dt}\big[ {\rm {\bf Re}}(\tilde{G}(t)){\rm {\bf Im}}(\tilde{G}(t))\big]\Big] $.
We have again the function $\tilde{G}(t)$ that is  explicity time-dependent
and then the time-ordering integrals cannot be converted to path-ordering
integrals. Therefore the integrals and the phases on the r.h.s. of
eq.(\ref{7}) are all time-dependent, as well as the functions with the overall
decreasing exponential.

In reference \cite{garrison88} Garrison and Wright present in eq.(4.9) the
value of the geometric  imaginary phase in the limit of  weak electric field and
$\gamma_b\gg \gamma_a$. In the weak electric field limit, the $SU(2)$ structure
in eq.(\ref{7}) disappears and the solutions of eqs.(\ref{6.1}) and (\ref{7})
become simpler,

\begin{subequations}\label{8}

\begin{equation}
{\rho}_{aa}(t) = \Big\{ {\rho}_{aa}(0) + i\int_0^t dt^\prime
(\frac{{\cal E}_0}{2} \vec{\mu}_{ab} \cdot  \vec{e} )
\Big[e^{i (\Delta  t^\prime - \phi(t^\prime))} {\rho}_{ba}(0) -
e^{-i (\Delta t^\prime - \phi(t^\prime))} {\rho}_{ab}(0)  \Big] e^{2\gamma t^\prime}
\Big\} e^{-2\gamma t} \label{8.1}
\end{equation}

\noindent and

\begin{equation}
\rho_{ab}(t) = \Big\{ {\rho}_{ab}(0) + i\int_0^t dt^\prime
( \frac{{\cal E}_0}{2}  \vec{\mu}_{ab} \cdot \vec{e})
\Big( 1 - 2 \rho_{aa}(0)\Big)e^{i (\Delta t^\prime - \phi(t^\prime))}
e^{(2i\Omega_+ + \gamma)t^\prime} \Big\} e^{- (2i(\Omega_+ + \omega_0)
+ \gamma)t}.  \label{8.2}
\end{equation}

\end{subequations}

\noindent Equations (\ref{8}) have terms with exponential decay
(imaginary factors) but they are time-dependent, as was obtained in
references \cite{physa2002,romero01}, in opposition to the results of
reference \cite{garrison88} (see eq.(\ref{gar1.2})).

Strictly speaking,  geometric phases appear in adiabatic processes that happen only 
in the limit  $T \rightarrow \infty$. However,  this limit is experimentally
implemented by taking $T \gg 2\pi/\omega$. The experimentally measured
path-dependent  phases are pretty much independent of the particular value
of $T$,
once the previous inequality is true. In eq.(4.3) of reference \cite{garrison88}
we have two imaginary phases:  one of them $(e^{-2Im(\beta_{-})})$
does not depend on the value of $T$ and consequently is path-dependent,
whereas the other one depends on the particular value of $T$. In our results
(eqs.(\ref{6.1}) and (\ref{7})), valid
for arbitrary intensities of the classical external electric field,
all the imaginary phases depend on the chosen value of $T$. It means that
the greater the period $T$ of  $\phi(t)$, the smaller the probability that the electron
be in the upper level; also, the smaller
the correlation $\rho_{ab}(t)$.

\vspace{.5cm}

In summary, we study the nature of the imaginary phase acquired by the
probability  of a two-level model coupled to an
external electric field with fixed frequency due to its interaction with the
electromagnetic vacuum. The interaction among the quantum system
(two-level system), the environment (vacuum) and the external classical electric
field is taken in the electric dipole aproximation and in the RWA.
Garrison and Wright in reference \cite{garrison88} discussed this
 same model using the non-hermitian Schr\"o\-dinger equation approach,
concluding that the probability  has an imaginary  phase that is path-dependent.
In this brief report we apply the matrix density formalism to study its
limit of $T\rightarrow \infty$ and from eqs.(\ref{6.1}) and (\ref{7}) we
conclude that all imaginary phases are time-dependent and
consequently depend on the chosen value of $T$. We are considering the
situation when the effects due to dissipation and 
time variation of Hamiltonian are of the same order and consequently
 the coherence of initial states are not preserved along
the whole period $T$. It is not surprising that those distinct approaches
(master equation and non-hermitian Hamiltonian) give
different results.

Finally we showed in the present work that even thought the non-hermitian
Hamiltonian method has been a very important and useful tool in
describing open systems, it should not be applied to discuss the
nature (time-dependent or path-dependent) of imaginary phases.


\section*{\bf Acknowledgements}

A.C. Aguiar Pinto thanks CNPq for financial support. M.T. Thomaz
thanks CNPq for partial financial support. The authors are in debt
with E.V. Corr\^ea Silva for the careful reading of the manuscript.



\begin{thebibliography}{99}

\bibitem{moiseyev98} N. Moiseyev, {\it Phys. Rep.} {\bf 302}
                     (1998) 212 and references therein.

\bibitem{baker83} H.C. Baker, {\it Phys. Rev. Lett.} {\bf 50} (1983) 1579.

\bibitem{baker84} H.C. Baker, {\it Phys. Rev.} {\bf A30} (1984) 773.

\bibitem{dattoli90b} G. Dattoli, A. Torre and R. Mignani,
                    {\it Phys. Rev.} {\bf A42} (1990) 1467.

\bibitem{themelis01} S.I. Themelis and C.A. Nicolaides, {\it J. Phys.}
                     {\bf B34} (2001) 2905.

\bibitem{dattoli88a} G. Dattoli, T. Hermsen, A. Renieri, A. Torre and
                     J.C. Gallardo,  {\it Phys. Rev.} {\bf A37} (1988) 4326.

\bibitem{dattoli88b} G. Dattoli, T. Hermsen, L. Mezi, A. Renieri and A. Torre,
                     {\it Phys. Rev.} {\bf A37} (1988) 4334.

\bibitem{berry84} M.V. Berry, {\it Proc. R. Soc.} {\bf A392} (1984) 45.

\bibitem{joye91} A. Joye, H. Kunz and Ch.-Ed. Pfister, {\it Ann. of Phys.} {\bf 208}
                 (1991) 299.

\bibitem {berry90} M.V. Berry, {\it Proc. R. Soc.} {\bf A430} (1990) 405.

\bibitem{zwanziger91} J.W. Zwanziger, S.P. Rucker and G.C. Chingas,
                      {\it Phys. Rev.} {\bf A43} (1991) 3232.
					 
\bibitem{uhlmann86} A. Uhlmann, {\it Rep. Math. Phys.} {\bf 24} (1986)
                  229.
			
\bibitem{uhlmann89} A. Uhlmann, {\it Ann. Physik Leipzig} {\bf 46} (1989)
                    63  

\bibitem{dabrowski1} L. Dabrowski and H. Grosse, {\it Lett. in Math.
                     Phys.} {\bf 19} (1990) 205.	
				  
\bibitem{dabrowski2} L. Dabrowski, {\it Il Nuovo Cimento} {\bf 106B}
                    (1991) 963.
				
\bibitem{anadan} Eric Sj\"oqvist {\it et al.}, {\it Phys. Rev. Lett.}
                 {\bf 85} (2000) 2845. 
				 
\bibitem{ericsson}  M. Ericsson {\it et al}., {\it Phys. Rev.} {\bf A67}
                   (2003) 020101.
				   
\bibitem{singh} K. Singh {\it et al}., {\it Phys. Rev.} {\bf A67} (2003)
                032106.
					  
\bibitem{garrison88} J.C. Garrison and E.M. Wright, {\it Phys. Lett.}
                     {\bf A128} (1988) 177.

\bibitem{dattoli90} G. Dattoli, R. Mignani and A. Torre, {\it J. Phys.}
                     {\bf A23} (1990) 5795.

\bibitem{mostafazadeh} A. Mostafazadeh, {\it Phys. Lett.} {\bf A264}(1999) 11.

\bibitem{korsch} H.J. Korsch and S. Mossmann, {\it Journ. of Phys.} {\bf A36}
                  (2003) 2139.

\bibitem{physa2002}  A.C. Aguiar Pinto,  K.M. Fonseca Romero and M.T. Thomaz,
                    {\it Physica} {\bf A311/1-2} (2002) 169. 

\bibitem{born28} M. Born und V. Fock, {\it Z. Phys.} {\bf 51} (1928) 165.

\bibitem{lamb87} W.E. Lamb, R.R. Schlicher and M.O. Scully, {\it Phys. Rev.}
                 {\bf A36} (1987) 2763.

\bibitem{louisell73} W.H. Louisell, {\it Quantum Statistical Properties
                     of Radiation}, John Wiley $\&$ Sons (N.Y. 1973).

\bibitem{agarwal73} G.S. Agarwal, {\it Phys. Rev.} {\bf A7} (1973) 1195.


\bibitem{carmichael93} H. Carmichael, {\it An Open Systems Approach to
                       Quantum Optics}, Lecture Notes in Physics {\bf m18},
                       Springer-Verlag (Berlin 1993).

\bibitem{romero01} K.M. Fonseca Romero, A.C. Aguiar Pinto and M.T. Thomaz,
                   {\it Physica} {\bf A307} (2002) 142. 
				   
\bibitem{time_operator} We define the time-ordering operator as

\vspace{-1cm}

\begin{eqnarray*}
 {\cal T} \Big( e^{\int_t^\tau {\bf A} (t^\prime) dt^\prime} \Big) & \equiv &
 {\bf 1} +  \int_t^\tau {\bf A} (t_1) dt_1
   +  \int_t^\tau {\bf A} (t_1) dt_1 \int_t^{t_1} {\bf A} (t_2) dt_2 + \nonumber \\
%
%
  & + & \int_t^\tau {\bf A} (t_1) dt_1 \int_t^{t_1} {\bf A} (t_2) dt_2
\int_t^{t_2} {\bf A} (t_3) dt_3 + \cdots  \nonumber
\end{eqnarray*}

\vspace{-0.5cm}

\noindent where $\tau$ can be smaller or bigger than $t$.




\end{thebibliography}
\end{document}